\documentclass[hidelinks,10pt,twocolumn]{article}

\usepackage{graphicx}
\usepackage[sectionbib]{natbib}
\usepackage{amsfonts}
\usepackage{hyperref}

\PassOptionsToPackage{hyphens}{url}\usepackage{hyperref}
\makeatother

\providecommand{\keywords}[1]{\textbf{\textbf{Keywords:}} #1}

\begin{document}

\title{A transparent referendum protocol with immutable proceedings and verifiable outcome for trustless networks}

\author{Maximilian Schiedermeier\footnote{{contributed also throughout previous employment at INSA Lyon - LIRIS.}}\\McGill University / DISL\&SEL, Canada\\\href{max.schiedermeier@mail.mcgill.ca}{maximilian.schiedermeier@mail.mcgill.ca}
\and Omar Hasan\\INSA Lyon / LIRIS, France\\\href{omar.hasan@insa-lyon.fr}{omar.hasan@insa-lyon.fr}
\and Tobias Mayer$^*$\\Verimi / Head of IT Architecture, Germany\\\href{tobias.mayer@verimi.com}{tobias.mayer@verimi.com}
\and Lionel Brunie\\INSA Lyon / LIRIS, France\\\href{lionel.brunie@insa-lyon.fr}{lionel.brunie@insa-lyon.fr}
\and Harald Kosch\\Unversit\"{a}t Passau, Germany\\\href{harald.kosch@uni-passau.de}{harald.kosch@uni-passau.de}}


%

\pagestyle{plain}
\setcounter{page}{1}
\pagenumbering{arabic}

\maketitle

\begin{abstract}
High voter turnout in elections and referendums is very desirable in order to ensure a robust democracy. Secure electronic voting is a vision for the future of elections and referendums. Such a system can counteract factors that hinder strong voter turnout such as the requirement of physical presence during limited hours at polling stations.
However, this vision brings transparency and confidentiality requirements that render the design of such solutions challenging. Specifically, the counting must be implemented in a reproducible way and the ballots of individual voters must remain concealed.
In this paper, we propose and evaluate a referendum protocol that  ensures transparency, confidentiality, and integrity, in trustless networks. The protocol is built by combining Secure Multi-Party Computation (SMPC) and Distributed Ledger or Blockchain technology. The persistence and immutability of the protocol communication allows verifiability of the referendum outcome on the client side. Voters therefore do not need to trust in third parties. We provide a formal description and conduct a thorough security evaluation of our proposal.
\end{abstract}

\keywords{
E-Voting, Trustless Networks, Political Networks, Transparency, Unlinkability, Blockchain}

\section{Introduction}
The voter turnout for the 2018 US midterm election was at $53.4\%$ \citep{elections}. Though compared to previous elections this is a high value, almost half of the population at voting age did not make use of their right to vote. While multiple reasons may lead to a decline, it is a common objective to render the voters' active participation as effortless and convenient as possible. A secure voting system based on remote clients could greatly improve the flexibility of potential voters. It would significantly reduce the administrative overhead of postal voting and eliminate voters' obligations to be physically present at a voting station during limited hours.\\
In this paper, we focus on referendums, which can be seen as a special instance of elections, with only two options offered for vote. Even though referendums are a simpler case of elections, implementing them correctly is still very challenging \citep{Clarkson2008} \citep{Springall2014}. Many parties may have an interest in manipulation of the outcome. Furthermore, we consider the context of trustless networks, where we assume that participants place little to no trust in one another and there does not exist a central trusted authority, or such an entity is not desirable. A breach of the ballot-secrecy may result in harmful consequences for voters. Given this sensitive context, voters naturally seek solutions they can trust. The classic analog way of conducting a secret referendum is having voters cast their ballots into boxes. This way they remain unlinkable to their votes. However, the logistic effort that is required for such an approach is tremendous. Ballot boxes must be set up, ballots with voting options must be printed and afterwards the counting must be realized by fair participants. The complex chain of implicit actions makes it hard to provide a proof of compliance for every single step. In this article we try to address this problem with an electronic-referendum scheme that puts emphasis on transparency that is to say, full client-sided verification of correctness.

\subsection{Contribution}
We propose a transparent referendum protocol with immutable proceedings and verifiable outcome. We define this immutability as the impossibility to tamper with the log of participant actions. Although there already exist protocols with similar ambitions, they commonly provide little evidence to the end user that the designated protocol was followed in practice. We suggest a protocol that is based on a creative combination of existing cryptographic tools. In order to achieve transparency, we also asses the viability of our proposal considering mobile clients and discuss to which extent the protocol can withstand adversary attacks. Our evaluation concentrates on confidentiality of votes, transparency and immutability of proceedings and a verifiable outcome.\\
The key idea behind our contribution is to use a blockchain as a complete log of all communication between participants. While the secrecy of individual votes is ensured by an SMPC scheme, the log allows anyone with access to the ledger, to autonomously compare the actual proceedings to the expected protocol. This verification can occur locally. Participants therefore gain proof of correctness by themselves and not via third parties.

\subsection{Outline}
In this article we first give a quick overview of related articles that pursue similar objectives. Some of them follow strategies that are very different to our approach. We point out the issues that they pose and how we intend to address them.\\
Next comes a brief presentation of our model, followed by an enumeration of the cryptographic tools we apply within our protocol. Afterwards, we delve into the exact phases and actions that describe our protocol, followed by an evaluation in two parts. The first part discusses how well our initial objectives are met by the proposal. The second provides a security analysis where we evaluate different adversary strategies and their potential impact. Finally we present our conclusions and delineate the potential topics of further investigations.

\section{Related Work}
\label{work}
In this section, we present articles that discuss how to design a protocol for electronic referendums. For each one, we outline the key idea and highlight associated disadvantages.

\bigskip
In \citep{Benaloh1987} the authors describe how secret sharing schemes can be used as SMPC for Secret-Ballot elections. This work unarguably is the cryptographic foundation to our proposal. However Benaloh's formal model by itself provides no practical transparency for the participants. In his approach, security lies entirely in the applied threshold system, that is to say, participants have no dynamic feedback on the effectiveness of the applied security mechanisms. Our proposal not only protects the privacy of voters, it also transparently monitors if ballots have been potentially compromised.

\bigskip
In \citep{Diaz2009}, an architecture for a privacy-aware electronic petition system is suggested and evaluated. As petitions typically express only two opinions (non-participation meaning approval, participation meaning disapproval to a topic), it can be considered a referendum system. The core element in this approach is involving anonymous certificates to elegantly restrict the referendum to eligible participants and eliminate double-spending in a privacy aware manner. 
	However the suggested protocol does not provide enough transparency for an anonymous voter's participation: 
	The act of participation by signing is not publicly transparent, therefore a dishonest petition server could discard signatures. The outcome would be indistinguishable from a case where the voter has never even contacted the server. Notably the voter has no way to prove the misbehavior of the sig\-nature-server.
	While our approach also involves anonymous credentials, we make sure that the semantic of issued tokens is independent of effectuated voting decisions. This allows us to ensure transparency, which ultimately renders dishonest server behavior detectable.
	
\bigskip
\citep{Zyskind} and \citep{Zyskind2015} provide a description of the Ledger-enabled SMPC platform (\textit{ENIGMA}). Our contribution differs in two aspects:	\begin{enumerate}
	\item{\textit{ENIGMA} was not explicitly designed for referendums. Though the authors mention a general compatibility for such scenarios, its applicability for this context is not assessed in much detail.}
	\item{In their platform, the ledger is neither an exclusive data-store, nor is it used as the exclusive channel for inter-participant communication. Therefore participants do not obtain the same level of communicative transparency as in our solution.}
\end{enumerate}
	
\bigskip
\citep{Cortier2019} rely on a threshold system that can defend the secrecy of ballots until a fixed number of colluding adversaries. However their protocol provides no control mechanism to monitor whether such collusion was attempted or has already occurred. As such voters can not obtain certainty  that their votes have actually remained undisclosed.

\bigskip
\citep{Bursuc2019} identified similar objectives. They introduce a metric to measure voter privacy and examine how compromised systems perform under that metric. In reaction to this evaluation they then suggest a protocol that performs well, given the metric. However, their protocol is very focused on that specific aspect and provides no mechanisms for other important goals, such as the prevention of ballot dropping.

\bigskip
Very related to our approach is a proposal by \citep{Li2019}, where an IoT enabled protocol is discussed. The presented approach gains security by persisting encrypted votes in a blockchain. However there are two fundamental differences to our approach:\begin{enumerate}
\item{It does not include a client side analysis of communication meta-data, excluding an additional verification of protocol proceedings.}
\item{In the described model, there is a clear and intentional separation between the blockchain infrastructure and the voting devices. For registration and notably casting of ballots, the voters access the blockchain via a gateway. This separation of blockchain and clients also eliminates the possibility to perform integrity checks on client's side. Clients thus have to rely on external entities for full integrity checks of the blockchain.}
\end{enumerate}

\bigskip
\citep{Lee2016} describe a blockchain based voting protocol. In contrast to our proposal, their solution involves a trusted third party for vote filtering.

\bigskip
\citep{Ayed2017} also suggest a blockchain based voting system. However, in their system the blockchain arranges persisted votes in an immutable order. Therefore, voters can not update their vote, once it has been submitted. Our system does not rely on such a mechanism and therefore does not come with this restriction.

\bigskip
In \citep{Riemann2015}, the authors introduce a taxonomy of further notions for distributed voting protocols.

\section{Our Model}

\subsection{Participants}
We distinguish between physical entities, identifiers and roles. Each physical entity possesses a unique and anonymous identifier. Furthermore, there are three roles that the physical entities can personify. A single entity can personify multiple roles, but not all combinations are allowed. The restrictions are explained in section \ref{roles}.

\subsubsection{Roles}
Our protocol involves the following roles:\label{roles}

\begin{itemize}
\item{\textbf{Initiator}: The initiator ensures all participants obtain the information required for the protocol execution.\\This role $I$ is represented by a single physical entity $init$. The initiator provides a referendum context that comprises all information required by other participants to follow the referendum procedure. It is the only action $init$ ever performs. He notably does not participate in the subsequent voting or counting. The physical entity behind $init$ must not personify another role. This restriction hinders collusion, as it isolates referendum preparations from the entities executing the protocol.}
\item{\textbf{Voters}: Voters are the devices of natural persons eligible to provide their opinion on the referendum context.\\We define the eligible set of $k$ physical voter entities to a given referendum as: $V = \{v_1, ..., v_k\}$.}
\item{\textbf{Workers}: Workers contribute to the execution of the protocol's underlying SMPC and provide intermediate results required to compute the referendum outcome and verification checksum.\\The set of $n$ physical worker entities is a subset of the voter entities:  $W = \{w_1, ..., w_n\}$, $W \subset V$. Workers are an example for physical entities personifying multiple roles. The physical entity behind each worker also, at some point acts as a voter. One advantage of this decision is that the total amount of entities, required to run our protocol decreases by $|W|$. In general, allowing a single entity to act on behalf of multiple roles is critical, as this gathers additional information at an entity. However, in this case the applied security mechanisms ensure that knowledge about a single ballot does not enable the worker to infer further information.}
\end{itemize}

\subsubsection{Identities}
\label{identities}
When we talk of \textit{participants} $P$, we implicitly mean the physical entities behind voters and workers.\footnote{Although with the definition $P = V \cup W$, $P$ is equal to $V$, we intentionally introduce $P$ for participants. By using $P$ instead of $V$, it becomes more clear that we are not only interested in voter behavior.} Participants do not know one another directly, but only by an anonymous pseudo-identifier $\bar{p}$. Likewise we introduce the set of all pseudo-identifiers as $\bar{P}$. Only for illustration purposes, we denote a mapping function $id:P\rightarrow\bar{P}$ that translates a specific entity $p \in P$ to its associated identifier $\bar{p} \in \bar{P}$. It is important to state that in practice no entity must ever possess such a function. Participant anonymity is an essential element in our protocol. From this point on when we talk of \textit{identifiers}, we implicitly mean \textit{pseudo-identifiers}.\\
Each participant holds a keypair. The private key is used for signatures and decryption. It never leaves the participant. The public key is used for encryption and also serves as a participant's identifier. We assume, that the initiator holds a complete list of all eligible voters' identifiers $\bar{V} = \{ id(v) | v \in V\}$.\\We consider this to be a fair assumption, since \textit{Diaz et al.} demonstrated how anonymous credentials can be issued among eligible voters, using an external credential server \citep{Diaz2009}. 

\subsection{Ledger}
A key component of our model is an immutable and integrity-protected data-store that is directly accessible by all participants. This is the ledger $L$. Access to the ledger enables the retrieval of persisted records and submission of new records. Persisted records however can be neither modified nor erased. 
\subsubsection{Ledger-restricted Communication}
Every participant locally operates a ledger-access node that allows him to retrieve records, submit new records and notably fully verify the ledger's integrity locally. We use the ledger as the \textit{exclusive} communication medium among participants. As participants only know one another by their identifiers, they exchange messages by adding and polling ledger records whenever they communicate.
\subsubsection{Message notation}
Every record added by communicating participants represents a message of format $m_{\alpha\beta}$. The index $\alpha$ specifies the sender's identifier, $\beta$ the recipient's identifier. In case of broadcast messages no recipient $\beta$ is provided. We distinguish between the following message types:
\begin{itemize}
\item{$b_\alpha$ with $\alpha = id(init)$\\The Initiator's broadcast message, specifying the referendum parameters.}
\item{$s_{\alpha\beta}$ with $\alpha=id(v_i), v_i \in V$, $\beta=id(w_j), w_j \in W$\\A voter sending a voting-related message to a worker.}
\item{$r_\alpha$ with $\alpha=id(w_j), w_j \in W$\\A worker's broadcast message that contributes to the referendum outcome.}
\item{$c_\alpha$ with $\alpha=id(w_j), w_j \in W$\\A worker's broadcast message that contributes to the referendum validation.}
\end{itemize}
The authenticity of message origins is ensured by the author's signature. As the registration of voters' public keys, described in \ref{identities}, can be realized over the ledger, it is fair to assume a trusted key-exchange among participants, prior to the referendum execution.

\subsection{Adversary Model}
We consider all voters and workers as potential adversaries. In section \ref{protoutline}, we outline the exact \textit{expected} behavior of referendum participants. Our adversary model covers that any Voter or Worker may deviate from this expected behavior at any time.
\subsubsection{Malicious communication}
In terms of message exchange, we consider:
\begin{itemize}
\item{submission of syntactically incorrect messages, for instance messages that lack mandatory meta-infor\-mation such as the signature.}
\item{submission of semantically incorrect messages. This notably covers the submission of values out of a legal range, as well as incorrect result-values for delegated computations. This may also arise  for header information, such as the sender field.}
\item{submission of messages that by format or content are not covered by the phase in progress.}
\item{inactivity where interaction is requested, that is deliberate non-communication.}
\end{itemize}
Adversaries may deviate from the expected behavior individually or in groups.

\subsubsection{Assumptions}
We assume that all information in $b_{id(init)}$, verifiable by each individual participant, is correct. This is a fair assumption, as the referendum will not take place unless the participants agree to the published parameters. Furthermore, we assume that it is infeasible for adversaries to fake RSA signatures or break encrypted messages. Adversaries are not able to resolve the physical identity of other participants by inspecting network traffic. This is realistic if participants use TOR. Finally, we assume that adversaries do not have the resources to break the ledger's integrity. We assume that the ledger is based on a blockchain thus this property is ensured. One of the characteristics of blockchains is that it requires a practically infeasible computational effort to break their integrity protection. \citep{Gaetani2015}\\
We assume that the protocol either results in a provably correct result, or the participants can detect anomalies. However, we do not expect the participants to correct detected issues.

\subsection{Objectives}
We set the following four objectives for our proposed protocol:
\begin{enumerate}
	\item{\textit{\textbf{Confidentiality}}: 
The referendum must be conducted in such a way that it is impossible to infer the choices made by individual voters.}
\item{\textit{\textbf{Transparency}}: 
The referendum must be \textit{transparent}. This means that every participant must obtain a complete trace of the operations performed, by whom and when. This notably covers the communication among participants throughout the referendum.}
\item{\textit{\textbf{Verifiability of the outcome}}: 
The referendum result must be verifiable to every participant. That means he must be able to autonomously evaluate the correctness of the result.}
\item{\textit{\textbf{Immutability of proceedings}}: 
Proceedings are the logs of all actions performed by participants from the moment of referendum initialization until the determination of the result. Proceedings must arise directly upon execution of the described actions. Once persisted, proceedings must be immutable. That is to say it must be impossibly to modify or even delete persisted proceedings.}
\end{enumerate}

\section{Building Blocks}
\subsection{Secret Sharing Scheme} 
Any secret sharing scheme supporting additive and multiplicative homorphic operations will serve for our protocol. We decided for the SEPIA \citep{Many2012} specification of Shamir's Secret Sharing, due to its good documentation and ease of integration. Shamir's Secret Sharing is an instance of SMPC schemes. As such, it allows to perform computations without having to reveal the original inputs to individual parties.

\subsubsection{t-n threshold systems}
A $t-n$ threshold system allows splitting a secret into $n$ shares in such a way that any $t$ of them suffice to reconstruct the original secret. Subsets with less than $t$ shares do not reveal any information about the secret. If the shares are distributed to multiple parties, we can thus create an effective mechanisms against collusion. If shares of a secret are distributed among $n$ parties, $t$ of them must cooperate, to reconstruct the secret. With a greater value $t$, the protection against collusive reconstruction rises. However, in case of a desired reconstruction, increasing the offset between $t$ and $n$ leads to enhanced robustness, as it makes the reconstruction redundant to the unavailability of single parties.\\The ratio of $t$ to $n$ can thus be adjusted, to meet a distributed protocol's security and robustness requirements.

\subsubsection{Homomorphic operations}
We make use of a secret sharing scheme that supports additive and multiplicative homomorphic operations. This means the secret sharing scheme provides a way to perform operations on the shares of different secrets, so that the fusion of the resulting shares provides values that are equivalent to calculations done on the original secrets. Shamir's secret sharing supports both additive and multiplicative homomorphic operations. However the multiplicative component has side effects that limit its practical application. Specifically, it increases the amount of shares required for a later reconstruction of the result value. This problem is was first mentioned in \citep{Benaloh1987}. The practical consequences for our protocol are discussed in section \ref{sec}.

\subsection{Distributed Ledger Technology}
Our protocol relies on a precise log of communication that cannot be tampered with. We therefore use Distributed Ledger Technology as the communication channel among participants. Specifically, using a blockchain ensures that manipulation of persisted data is computationally infeasible. To do so an adversary would have to outperform the honest majority of mining   participants. 

\subsection{Asymmetric Encryption}
Although our protocol requires a complete listing of communication meta-data, there are good reasons to delimit the content of messages to the recipient. We therefore use asymmetric encryption to generate public-private keypairs which allow encryption and decryption of directed messages. Furthermore, these keys are used for message signing and authorship validation.

\subsection{Anonymous Credentials}
Anonymous credentials allow a restricton of services to specific users, without a need to verify identities at the moment of access. The key idea is to introduce an external entity that hands out cryptographic tokens to eligible users \citep{Chaum1985}. Those users can later use their credentials to gain admission to an access controlled service. Though modern implementations \citep{Lysyanskaya2001} respond to advanced requirements such as detection of double spending or a privacy aware verification of user specific attributes, we only make use of the key feature, as it allows the anonymous registration of eligible voters.

\section{The Protocol} 
\begin{figure}
\centering
\scalebox{.22}{
\includegraphics{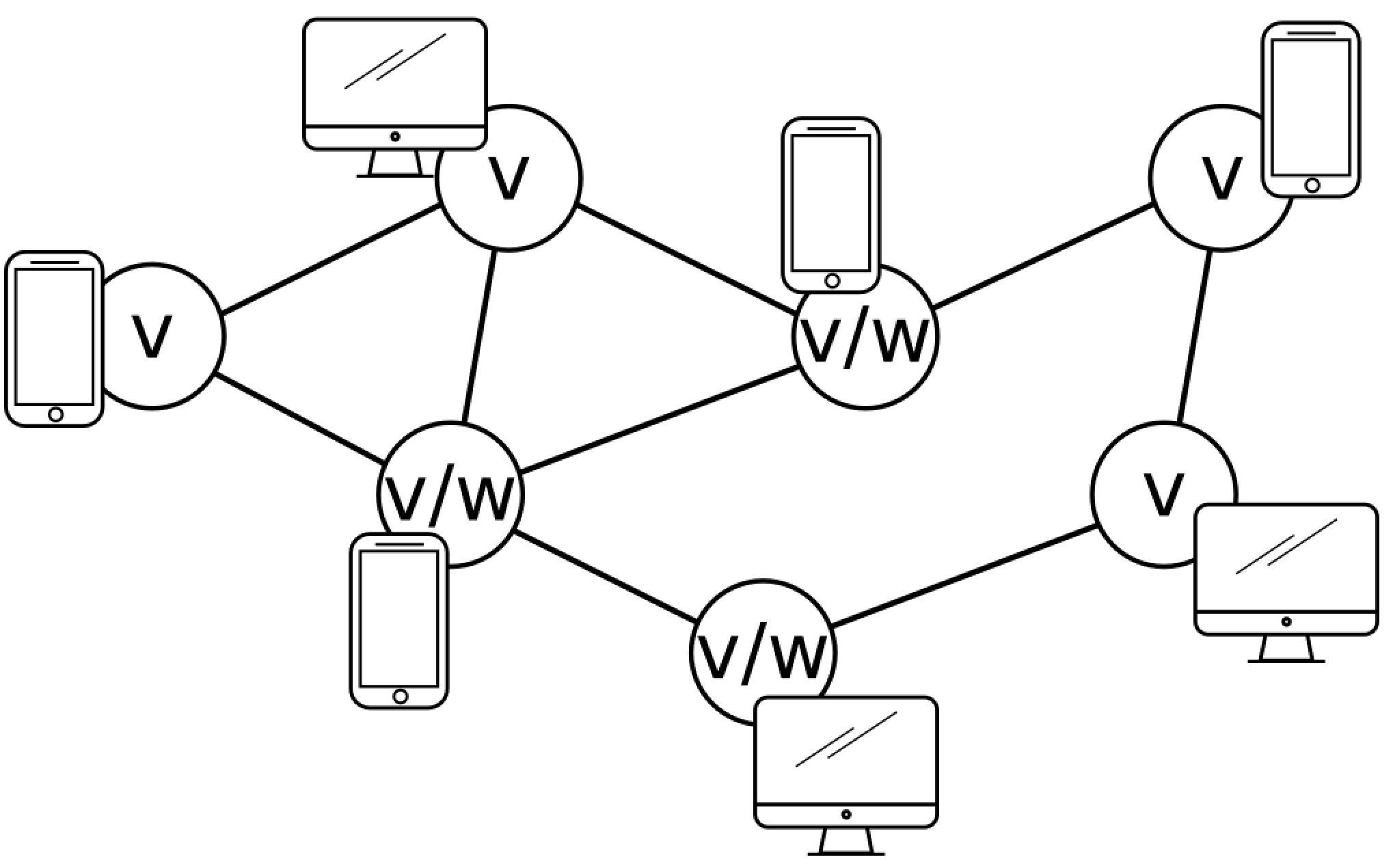}}
\caption{Illustration of \textit{referendum participants connected as blockchain nodes.
\label{p2p}} Every referendum participant replicates the ledger. Although the nodes constantly synchronize, referendum related messages are exchanged exclusively via ledger-records. Therefore all clients hold a transparent copy of the proceedings. As pictured above, the protocol does not bind specific roles to particular hardware.}
\end{figure}

The key idea behind our contribution is to use the ledger as a complete log of all communication between participants. This allows anyone, with access to the ledger, to autonomously compare the actual proceedings to the expected protocol. This verification can occur locally. Participants therefore gain proof of correctness by themselves and not via third parties.\\Furthermore, the anonymity of individual participants effectively prevents communications via side channels. This is discussed in more detail in section \ref{sec}.\\
As our protocol is based on a secret sharing system, the introduction of a public ledger is counter-intuitive. Secret sharing systems usually gain security by dividing information into separate shares. Yet we suggest to store such shares side by side in a public ledger. We make this design feasible, by additionally protecting persisted shares with asymmetric encryption. This ensures that only an intended target entity has access to a specific set of sensitive information. At the same time, the ledger as an exclusive communication channel allows us to monitor the message meta-data of all participants. This allows clients to autonomously verify the absence of adversary collusion, targeted on the underlying $t-n$ threshold system. Secret communication via side-channels is not an issue, as participants only know another by their anonymous identity.\\
As the verification of the ledger's integrity by itself does not require clients to actively mine, we consider it reasonable to enable mobile clients as blockchain replicating nodes. This is a valid assumption, given two conditions:
\begin{enumerate}
\item{The used blockchain serves exclusively for the purpose of the current referendum. By restricting the ledger content to this specific payload, the blockchain's data volume is significantly reduced. This is an essential decision, since popular public chains can easily exceed the storage capacity of a mobile client and therefore render a replication infeasible.}
\item{As shown in Figure \ref{p2p}, a portion of the participants relies on non-mobile hardware, bearing the resources for active mining. This is likewise an important condition, as the blockchain only provides integrity protection when an honest majority of miners can not be computationally outperformed by adversaries \citep{Gaetani2015}.}
\end{enumerate}

\subsection{Protocol Overview}
Our referendum protocol is based on a secure multi-party computation scheme, with the restriction added that all inter-participant communication occurs exclusively over a public ledger. That is to say, parties can only communicate by placing public messages in the ledger. Messages clearly state the recipient and are furthermore signed by the author. This provides a transparent and clear trace of all arising inter-participant communication.\\The SMPC by itself allows a privacy aware computation of the referendum outcome. The SMPC's homomorphism ensures that computing entities do not learn about sensitive input data, since they work on an encrypted transformation of the data.\\Proof of conformity to the designated protocol is supported by the ledger's immutability. Voters can analyze communication meta-data of the executed SMPC. This way every participant can assess whether the actual communication followed the protocol. As all information required to perform this validation is stored in the ledger, referendum participants can implement all compliance checks locally, without the need to trust third parties. This allows the protocol to function in trustless network environment.\\
Ultimately, after a successful validation of the proceedings, each voter holds the certainty that the outcome was determined correctly and no vote has been compromised.

\subsection{Protocol Outline}
\label{protoutline}
\begin{figure}
\centering
\scalebox{.18}{
\includegraphics{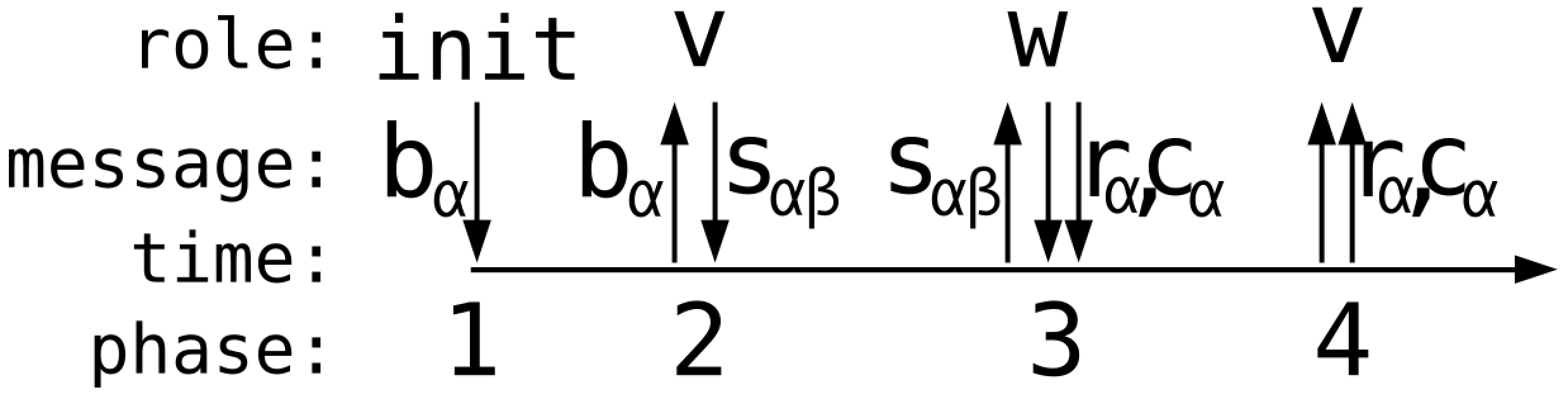}}
\caption{Illustration of \textit{protocol phases.
\label{phases}} Downward arrows indicate the persistence of messages \textit{types} into the ledger, upward arrows indicate the lookup of messages (indicated by \textit{type}). Time advances from left to right.}
\end{figure}

Figure \ref{phases} illustrates how individual roles chronologically submit and retrieve messages to the ledger. For each action, it also indicates the corresponding protocol phase. 
\begin{enumerate}
\item{\textit{\textbf{Initiation}}: In this step, the referendum conditions are written to the ledger: \textit{Referendum context, voting options, identities of registered voters, etc}...}
\item{\textit{\textbf{Vote submission}}: Voters look up the referendum conditions and deposit their ballots, secured by the secret sharing scheme and asymmetric encryption. }
\item{\textit{\textbf{Intermediate result computation}}: Workers perform homomorphic operations on the secured ballots, then write intermediate results and checksums back to the ledger.}
\item{\textit{\textbf{Determination and validation of the outcome}}: Voters pick up the intermediate results and checksums to determine the final outcome and run verifications.}
\end{enumerate}
~
The next section provides more details regarding the individual phases.

\subsubsection{Initiation}
The goal of the first phase is to ensure that all participants operate on identical referendum parameters. The referendum initiator $init$ ensures this with a single broadcast message:
\begin{enumerate}
\item{$init$ places an initial broadcast message $b_{id(init)}$ in the ledger. The content of this message, $\tilde{b}_{id(init)}$  accumulates all static referendum parameters. It includes:
\begin{itemize}
\item{The identities (public keys) of all eligible voters: $\bar{V} = \{id(v_i) | v_i \in V\}$.}
\item{A subset of identities that names the designated workers: $\bar{W} = \{id(w_j) | w_j \in W \}$ as well as the individual share affiliation. The latter is required by the voters in the next phase, so they know which share belongs to which worker.} 
\item{The referendum context and semantics of numeric voting options. This can be for instance:\\ \textit{Are cats cooler than dogs? Yes = $+1$, No = $-1$}.}
\item{A set of time-stamps that define the transitions between subsequent phases $Q = \{q_{1-2}, q_{2-3}, q_{3-4}\}$. The fixed time stamps are required to ensure that at the start of each phase all required input data is present in the ledger. As $q_{1-2}$ marks the transition to phase 2, this timestamp matches the moment of placing $b_{id(init)}$ in the ledger.}
\end{itemize}
By communicating these conditions through a ledger, all participants obtain the exact same understanding of the expected referendum proceedings. This initial message contains all information required to outline further communication among participants.}
\end{enumerate} 

\subsubsection{Vote Submission}
In the second phase, voters cast their votes. Each voter $v_i \in V$ does the following:
\begin{enumerate}
\item{$v_i$ retrieves the initiator's broadcast message from the ledger.}
\item{$v_i$ secretly chooses his personally preferred voting option and determines the corresponding numeric value $\psi_i$. The mapping is specified in $b_{id(init)}$.}
\item{Based on $\psi_i$, voter $v_i$ then generates a set of $n$ shares $\{\sigma_{i1}, ..., \sigma_{in}\}$. He does so  following a $t-n$ threshold secret sharing scheme. The exact parameters for this step are provided in $b_{id(init)}$.}
\item{Each generated share is intended for a specific worker $w_j$. Voter $v_i$ encrypts each generated share $\sigma_{ij}$ with the corresponding worker $w_j$'s public key $\tilde{w}_j$. The exact mapping of shares to workers is once more described in $b_{id(init)}$. The target worker's id is also the public key to use for encryption.}
\item{$v_i$ packs all $n$ cypher-shares $\tilde{s}_{ij} =pub_j(\sigma_{ij})$, $j \in \{1, ..., n\}$ individually into  n messages ${s_{ij}}$ and initiates their persistence in the ledger. The horizontal arrows in Figure \ref{ledger} illustrate this step.}
\end{enumerate}
Voters can perform the above steps until timestamp $q_{2-3}$ is reached. Repeated submissions before the deadline are allowed. Those are considered an update to one's own ballot. Messages ${s_{ij}}$ submitted after $q_{2-3}$ are considered non-compliant to the protocol and will be ignored.

\subsubsection{Intermediate result computation}
\begin{figure}
\centering
\scalebox{.23}{
\includegraphics{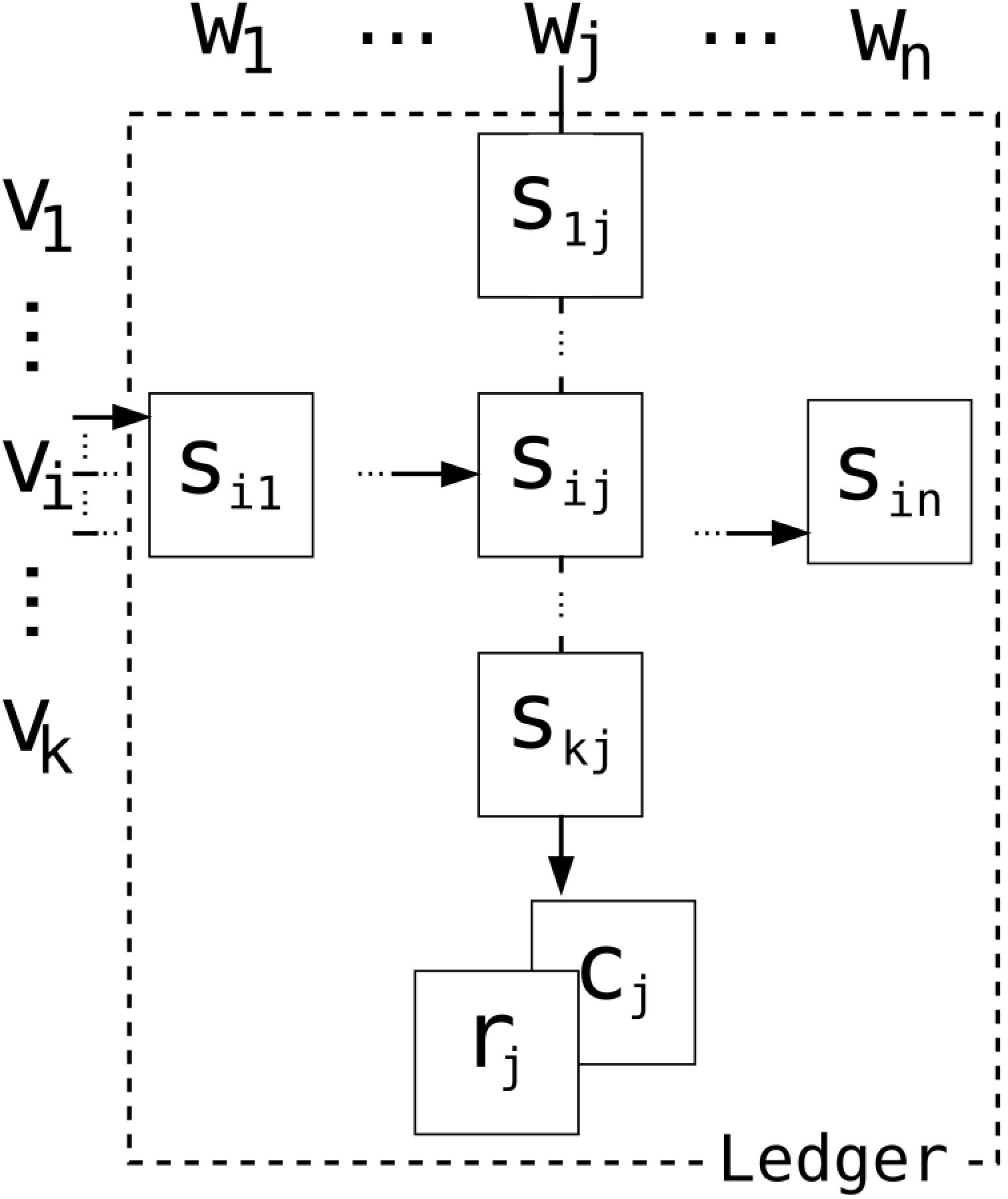}}
\caption{Illustration of \textit{vote submission} by a voter $v_i$ and \textit{intermediate result computation} by a worker $w_j$. Note that all messages arising throughout these steps are persisted in the ledger.
\label{ledger}}
\end{figure}

In the third phase, each worker $w_j$  performs the following actions to contribute intermediate result values for the referendum outcome and checksum computations:
\begin{enumerate}
\item{$w_j$ retrieves the set of $k$ encrypted share-messages destined to him: $\{s_{1j}, ..., s_{kj}\}$.}
\item{$w_j$ retrieves the payload of received messages and this way holds $k$ shares, each encrypted with his public key: ${\tilde{s}_{1j}, ..., \tilde{s}_{kj}}$.}
\item{$w_j$ decrypts every single share using his private key and obtains a set of $k$ unencrypted shares: $\{\sigma_{1j}, ..., \sigma_{kj}\}$. These are the $k$ shares, the voters $V=\{v_1, ..., v_k\}$ securely communicated to him via ledger.}
\item{Based on $\{\sigma_{1j}, ..., \sigma_{kj}\}$, $w_j$ participates in the homomorphic calculation of intermediate result shares:
\begin{itemize}
\item{He contributes to obtaining the sum of all votes, with an intermediate result share $\tilde{r}_j$}.
\item{He contributes to obtaining the sum of all squared votes, with an intermediate result share $\tilde{c}_j$}.
\end{itemize}
The sum of squared votes will later serve to detect illegal inputs. Note that intermediate result shares $\tilde{r}_j, j \in \bar{W}$, respectively $\tilde{c}_j, j \in \bar{W}$ must be combined to obtain the actual results.}
\item{$w_j$ converts $\tilde{r}_j$ and $\tilde{c}_j$ to broadcast messages $r_j$, $c_j$ and makes those get persisted in the ledger.}
\end{enumerate}
The execution of the above steps by a worker $w_j$, leading to persistence of $r_j$ and $c_j$, is illustrated in Figure \ref{ledger} by a downward arrow.\\
$q_{3-4}$ marks the moment by which workers must have their intermediate results persisted.

\subsubsection{Determination and validation of the outcome} 
In the final phase, voters individually reconstruct the referendum outcome and evaluate public proceedings' conformity. To achieve this, every voter $v_i$ performs the following actions on the intermediate result shares $\{\tilde{r}_j | j \in \bar{W}\}$ and $\{\tilde{c}_j | j \in \bar{W}\}$:
\begin{enumerate}
\item{$v_i$ picks up the corresponding result and checksum messages:  $\{r_j | j \in \bar{W}\}$ and  $\{c_j | j \in \bar{W}\}$.}
\item{$v_i$ obtains two sets of shares, by combining the message payloads: $\{\tilde{r}_j | j \in \bar{W}\}$ and  $\{\tilde{c}_j | j \in \bar{W}\}$}
\item{He removes the protection of the threshold system for two specific values. Precisely, he combines the intermediate result shares $\{\tilde{r}_j | j \in \bar{W}\}$, respectively  $\{\tilde{c}_j | j \in \bar{W}\}$. These sets of shares express the homomorphic equivalent of:
\begin{itemize}
\item{The referendum outcome, $r = \sum_{i \in V}\psi_i$}
\item{A referendum checksum, $c = \sum_{i \in V}\psi_i^2$}
\end{itemize}
Consequently by combining the corresponding shares, $v_i$ obtains $r$ and $c$. The checksum $c$ allows the detection of illegal votes. As all votes are expected to be either of $\pm 1$, it must hold that $c = k$. If that is not given, the participant directly knows that at least one illegal input value was submitted.\footnote{Still, it is possible to generate a valid checksum with cleverly arranged illegal input values. We discuss this threat in section \ref{sec}.}}
\end{enumerate}

\newpage

\section{Analysis of objective fulfillment} 
\label{eval}
In this section we evaluate how well the individual objectives are met by the suggested protocol.
\subsection{Immutability of the referendum proceedings}
Proceedings are immutable whenever they are preserved in a way that renders retroactive tampering infeasible. Given the presented protocol, proceedings can be expressed by a complete log of  partici\-pant-exchanged messages. As those messages are exchanged publicly through the ledger, the ledger content itself serves as complete transcript of referendum proceedings. We ensure the ledgers exclusive status as targeted communication medium by concealing the physical identity of participants behind pseudonyms. Since the blockchain ensures the immutability of persisted records, we obtain an immutable log of the referendum proceedings.
\subsection{Confidentiality of votes}
A ballot is secret if no entity other than the voter himself knows the submitted value. Our protocol applies a strong protection of votes, by first splitting them according a secret sharing scheme and then encrypting the obtained shares asymmetrically. Unless an adversary manages to break asymmetric encryption or secretly gather the private keys of $t$ workers for a collusive ballot reconstruction, the confidentiality of submitted votes remains ensured. Though asymmetric encryption mechanisms are theoretically breakable, it is commonly assumed a computationally infeasible task. That is to say with current hardware it is extremely unlikely for an adversary to reconstruct a secret without the required key material.\\
Furthermore, by analyzing the ledger, voters can reconstruct the message flow among participants and exclude even the possibility that workers colluded to reunite shares. As workers only know another by their pseudo-identifiers, they can not secretly establish a communication side channel for collusion.
\subsection{Referendum validation}
To verify the correctness of the referendum outcome, each participant must be able to validate that two conditions are met:
\begin{enumerate}
\item{The inputs that the outcome evaluation occurred on, are valid. This means all votes must be valid numeric options. As we will see in section \ref{sec}, this condition restricts the range of valid parameters for the $t-n$ threshold system.}
\item{The evaluation itself was conducted correctly. This means that the intermediate results computed by the workers must be correct for the provided inputs.}
\end{enumerate}
The second condition can be ensured by redundancy. The polynomial based secret sharing scheme allows to detect and ignore outliers. Imagine 10 sampling points are provided for a polynomial of degree two. Now, if nine of them match the polynomial but a single point does not, this would suggest that the 10th support is incorrect.\\
Assuming that intermediate results are verifiable, the worker-provided checksum allows a privacy aware input validation. 
\subsection{Transparency}
A referendum is considered transparent if all participants possess a correct and complete log of all actions performed throughout the entire referendum. In our model, all actions eventually result in communication. As we force all communication to run through the ledger, the trace of deposited messages provides a transparent and verifiable log of actions.

\section{Security analysis}
\label{sec}
In this section, we evaluate whether adversary strategies are detrimental to the suggested protocol:
\begin{itemize}
\item{\textbf{Intentional inactivity}: Any participant can violate the protocol by intentional inactivity where interaction is expected. Eligible voters can choose not to distribute shares or only send them to a subset of workers. A worker can ignore the expected submission of intermediate result shares.\\Although the payload of vote-messages is encrypted, all participants can inspect the ledger content and detect if eligible voters are inactive or do not communicate with all designated workers. The default strategy is to systematically ignore all vote-shares of voters that do not comply to the expected behavior. This way the disadvantage of inactivity lies entirely with the adversary. Inactive workers are harder to prevent, but the redundancy of the t-n-threshold system allows a determination of the evaluation \textit{outcome} until up to $n-t$ inactive workers. However, in terms of the referendum outcome's \textit{checksum}, the boost of sampling points required for reconstruction, lowers the protocol's robustness to a tolerance of at most of $n-t^2$ inactive workers.  \citep{Benaloh1987}}
\item{\textbf{Syntactically incorrect messages}: Participants can violate the protocol by sending syntactically incorrect messages.\\Syntactic errors can be easily detected with syntax-schemes. The default strategy is to ignore any syntactically incorrect message. This way, messages that are in no relation to the protocol also have no impact. If ignoring the message results in an interpreted participant inactivity, the above inactivity analysis is applicable here, too.}
\item{\textbf{Impersonation}: Participants may try to illegally send messages in the name of another participant.\\Impersonated messages are easy to detect, since their signature does not match the expected author. Messages with invalid signature are systematically ignored.}
\item{\textbf{Invalid voting options}: Voters are expected to vote for either $\pm 1$. However, as their shares are submitted in encrypted form, they might try to boost their influence with higher (or lower) numeric values.\\For colluding participants, it is possible to arrange invalid votes in a way that the input validation checksum is still fulfilled.\footnote{Example: Imagine two votes $\psi_1 = \sqrt[2]{1.5}, \psi_2 = \sqrt[2]{0.5}$ are submitted, their checksum is $\psi_1^2 + \psi_2^2 = 2$, while the resulting vote impact is $\psi_1 + \psi_2 \neq 2$.} However, this attack is not in the interest of the adversaries, since it can only diminish the overall influence of the outcome. If parties collusively submit illegal inputs that pass the validation, the impact of those inputs is lower than the impact they would have achieved with legal input values. This is a consequence of the \textit{Cauchy-H{\"o}lder inequality}: $\sum_{k=1}^n | x_k y_k | \leq (\sum_{k=1}^n {{| x_k|}^p)}^{1/p} (\sum_{k=1}^n {{| y_k|}^q)}^{1/q}$, with $n \in \mathbb{N}, \{x_1, ..., x_n\}, \{y_1, ..., y_n\} \in \mathbb{R}, p, q \in [1, \infty )$.\footnote{If we chose $p=2, q=2, y_k=1$, the inequality is reduced to $\sum_{k=1}^n | x_k | \leq \sqrt[2]{\sum_{k=1}^n {|  x_k | ^2 }} \sqrt[2]{n}$. However, the client side checksum verification ensures that $ \sum_{k=1}^n { | x_k| } = n$, which further reduces the inequality to $\sum_{k=1}^n { | x_k | } \leq n $. This maximum value is obtained with valid inputs $x_k \in \{ -1, +1\}$, rendering a collusive construction of illegal inputs pointless, since such inputs cannot surpass the impact of valid values, on the referendum.}\\Furthermore, it is not a severe threat to the protocol as it  requires collusion.} 
\item{\textbf{Incorrect intermediate results}: Workers might submit incorrect intermediate results on purpose. In case of an extreme threshold system configuration with $t=n$, the existence of incorrect result shares is neither detectable nor correctable. However, with rising share redundancy, an honest majority of workers can push incorrect shares into a detectable outlier position (see section \ref{eval}). However, massively colluding adversaries could also push an honest minority into an outlier position. Another inconvenience for worker adversaries is that they can not predict the effects of their manipulation. Given the SMPC, an altered value can influence the result in either direction. As discussed previously, we furthermore hinder collusive attacks by concealing the participants' natural identities.}
\item{\textbf{Double voting}: Voters can repeat the generation, encryption and distribution of shares.\\As the encrypted vote-shares are exchanged via the public ledger and sender and recipient remain un-encrypted header attributes, double voting is easily detectable. The default strategy is to discard all but the most recent share that a specific voter submits to a specific worker. A voter can thus update her choice, but not increase the impact.}
\item{\textbf{De-anonymization}: Participants might be interested to identify the physical entity that operates behind a participant pseudonym. This would enable outside-ledger undetected communication. As all network traffic runs over TOR connections, a de-anonymization is not feasible.}
\item{\textbf{Communication side channel creation}: Adversaries may try to secretly establish an alternate platform for direct communication, parallel to the ledger.\\Though secret inter-participant communication is a severe threat to the protocol's transparency and opens a gate for further attacks, it is not trivial to establish. A resilient system can counter this by setting the threshold-value reasonably high. Specifically, this means that the probability of the random workers to fall into societal cliques must be minimized. If adversaries do not already know their physical identities, they have to communicate publicly, as they do not know who to address to. Adversaries publicly declaring their will to collude can be easily detected.}
\item{\textbf{Voter exclusion}: In section \ref{work}, we criticized the usage of an anonymous credential server. However, in our case anonymous credentials are only used for registration, not for voting. In \citep{Diaz2009}, a voter cannot expose a dishonest behavior of a petition server. He cannot prove his previous interaction with the server and it would reveal his voting decision. In our case both does not apply. The registration itself can be logged in the ledger. Likewise the keys of registered voters, since they can be logged as part of the public init message, $b_{id(init)}$. Thus a legitimate voter could easily prove his exclusion by a malicious server.}
\end{itemize}

\section{Conclusion}
\subsection{Objectives fulfillment} 
Our work demonstrates how the challenges of electronic referendums can be answered with a creative combination of existing approaches. By bringing together the potential of blockchain technology and secure multiparty computation, we constructed a highly transparent referendum protocol that allows participants to autonomously verify proceedings and outcome. Traditional $t-n$ threshold based systems gain security exclusively by selection of parameters that render successful collusive attacks unlikely. To the best of our knowledge there is no other system that further enforces security, by considering proofs that inspect communication meta-data, protected by ledger technology. This concept generates trust at client side, because with exception to the anonymous credential issuer a need for trusted third parties is eliminated.\\
We provided a realistic adversary model and analyzed how our protocol withstands corresponding attacks.
\label{futwork}
\subsection{Future Work}
In future research we would like to further investigate a meaningful selection of referendum parameters. We could also imagine to experiment with machine learning approaches, to reach an optimal trade-off between security and scaling. Also we would like to explore other input validation methods that have less impact on the voter-worker ratio. Another open question is how to best select the worker subset. Given the focus of our current work on the security aspects of the protocol, we are interested in performance evaluations of a practical implementation of the protocol, particularly in a mobile scenario.
\newpage

\bibliography{Bibliography}

\begin{thebibliography}{17}
\providecommand{\natexlab}[1]{#1}
\providecommand{\url}[1]{\texttt{#1}}
\expandafter\ifx\csname urlstyle\endcsname\relax
  \providecommand{\doi}[1]{doi: #1}\else
  \providecommand{\doi}{doi: \begingroup \urlstyle{rm}\Url}\fi

\bibitem[Ayed(2017)]{Ayed2017}
Ahmed~Ben Ayed.
\newblock {A conceptual secure blockchain-based electronic voting system}.
\newblock 9\penalty0 (3):\penalty0 1--9, 2017.

\bibitem[Benaloh(1987)]{Benaloh1987}
Josh~Cohen Benaloh.
\newblock {Secret sharing homomorphisms: Keeping shares of a secret secret
  (Extended Abstract)}.
\newblock \emph{Lecture Notes in Computer Science (including subseries Lecture
  Notes in Artificial Intelligence and Lecture Notes in Bioinformatics)}, 263
  LNCS:\penalty0 251--260, 1987.
\newblock ISSN 16113349.

\bibitem[Bureau(2019)]{elections}
U.S.~Census Bureau.
\newblock {US elections} voter turnout statistics.
\newblock
  \url{https://www.census.gov/library/stories/2019/04/behind-2018-united-states-midterm-election-turnout.html},
  2019.

\bibitem[Bursuc et~al.(2019)Bursuc, Dragan, Kremer, Bursuc, Dragan, Kremer,
  Bursuc, Est, Kremer, and Est]{Bursuc2019}
Sergiu Bursuc, Constantin-catalin Dragan, Steve Kremer, Sergiu Bursuc,
  Constantin-catalin Dragan, Steve Kremer, Sergiu Bursuc, Inria Nancy-grand
  Est, Steve Kremer, and Inria Nancy-grand Est.
\newblock {HAL Id : hal-02099434 Private votes on untrusted platforms : models
  , attacks and provable scheme}.
\newblock 2019.

\bibitem[Chaum(1985)]{Chaum1985}
David Chaum.
\newblock {Security without identification: Transaction systems to make big
  brother obsolete}.
\newblock 28\penalty0 (70), 1985.

\bibitem[Clarkson et~al.(2008)Clarkson, Myers, Clarkson, and
  Myers]{Clarkson2008}
Michael~R Clarkson, Andrew~C Myers, Michael~R Clarkson, and Andrew~C Myers.
\newblock {Civitas : Toward a Secure Voting System Civitas : Toward a Secure
  Voting System}.
\newblock 7875, 2008.

\bibitem[Cortier et~al.(2019)Cortier, Gaudry, Glondu, Cortier, Gaudry, and
  Glondu]{Cortier2019}
V{\'{e}}ronique Cortier, Pierrick Gaudry, Stephane Glondu, V{\'{e}}ronique
  Cortier, Pierrick Gaudry, and Stephane Glondu.
\newblock {Belenios : a simple private and verifiable electronic voting system
  To cite this version : HAL Id : hal-02066930}.
\newblock 2019.

\bibitem[Diaz et~al.(2009)Diaz, Kosta, Dekeyser, Kohlweiss, and
  Nigusse]{Diaz2009}
Claudia Diaz, Eleni Kosta, Hannelore Dekeyser, Markulf Kohlweiss, and Girma
  Nigusse.
\newblock {Privacy preserving electronic petitions}.
\newblock \penalty0 (2008):\penalty0 203--219, 2009.

\bibitem[Gaetani et~al.(2015)Gaetani, Aniello, Baldoni, Lombardi, Margheri, and
  Sassone]{Gaetani2015}
Edoardo Gaetani, Leonardo Aniello, Roberto Baldoni, Federico Lombardi, Andrea
  Margheri, and Vladimiro Sassone.
\newblock {Blockchain-based Database to Ensure Data Integrity in Cloud
  Computing Environments}.
\newblock 2015.

\bibitem[Lee et~al.(2016)Lee, James, Kim, and James]{Lee2016}
Kibin Lee, Joshua~I James, Hyoung~J Kim, and Joshua~I James.
\newblock {Electronic Voting Service Using Block-Chain}.
\newblock 11\penalty0 (2), 2016.

\bibitem[Li et~al.(2019)Li, Susilo, Yang, Yu, Liu, and Guizani]{Li2019}
Yannan Li, Willy Susilo, Guomin Yang, Yong Yu, Dongxi Liu, and Mohsen Guizani.
\newblock {A Blockchain-based Self-tallying Voting Scheme in Decentralized
  IoT}.
\newblock 2019.

\bibitem[Lysyanskaya and Camenish(2001)]{Lysyanskaya2001}
Anna Lysyanskaya and Jan Camenish.
\newblock {An Efficient System for Non-transferable Anonymous Credentials with
  Optional Anonymity Revocation}.
\newblock \emph{EUROCRYPT 2001}, 2001.

\bibitem[Many et~al.(2012)Many, Burkhart, and Dimitropoulos]{Many2012}
Dilip Many, Martin Burkhart, and Xenofontas Dimitropoulos.
\newblock {Fast Private Set Operations with SEPIA}.
\newblock \emph{Technical report, Department of Computer Science, ETH Zurich},
  2012.

\bibitem[Riemann and Grumbach(2015)]{Riemann2015}
Robert Riemann and St{\'{e}}phane Grumbach.
\newblock {Distributed Protocols at the Rescue for Trustworthy Online Voting}.
\newblock 2015.

\bibitem[Springall et~al.(2014)Springall, Finkenauer, Durumeric, Kitcat,
  Hursti, Macalpine, and Halderman]{Springall2014}
Drew Springall, Travis Finkenauer, Zakir Durumeric, Jason Kitcat, Harri Hursti,
  Margaret Macalpine, and J~Alex Halderman.
\newblock {Security Analysis of the Estonian Internet Voting System}.
\newblock \penalty0 (May):\penalty0 703--715, 2014.

\bibitem[Zyskind(2015)]{Zyskind}
Guy Zyskind.
\newblock {Enigma : Decentralized Computation Platform with Guaranteed
  Privacy}.
\newblock pages 1--14, 2015.

\bibitem[Zyskind et~al.(2015)Zyskind, Nathan, and Pentland]{Zyskind2015}
Guy Zyskind, Oz~Nathan, and Alex~Sandy Pentland.
\newblock {Decentralizing privacy: Using blockchain to protect personal data}.
\newblock \emph{Proceedings - 2015 IEEE Security and Privacy Workshops, SPW
  2015}, pages 180--184, 2015.
\newblock ISSN 9781479999330.

\end{thebibliography}

\end{document}